\documentclass[runningheads]{llncs}
\usepackage{cite}
\usepackage{amsmath,amssymb,amsfonts}
\usepackage{hyperref}
\usepackage{graphicx}
\usepackage{textcomp}
\usepackage{multicol}
\usepackage{multirow}
\usepackage{subfigure}
\usepackage{xcolor}
\usepackage{lscape} 
\usepackage{colortbl}
\usepackage{booktabs}
\usepackage{adjustbox}
\usepackage{orcidlink}

\title{Automated angiography analysis for multivessel Coronary Artery Segmentation and Stenosis Localisation with Limited Data}
\title{Automated Coronary Angiography Analysis for Multivessel Segmentation and Disease Classification with Limited Data}
\title{Multivessel Coronary Artery Segmentation and Disease Localisation using Deep Ensemble Learning}
\title{Multivessel Coronary Artery Segmentation and Stenosis Localisation using Ensemble Learning}

\begin{document}

\author{Muhammad Bilal\inst{1}$^,$\inst{2}$^,$\inst{*} \and Dinis Martinho\inst{3} \and Reiner Sim\inst{4} \and Adnan Qayyum\inst{5}$^,$\inst{6} \and Hunaid Vohra\inst{7} \and Massimo Caputo\inst{7} \and Taofeek Akinosho\inst{2} \and Sofiat Abioye\inst{2} \and \\ Zaheer Khan\inst{2} \and Waleed Niaz\inst{2} \and Junaid Qadir\inst{5}}

\institute{Birmingham City University, Birmingham, United Kingdom \\ \and
University of the West of England (UWE), Bristol, United Kingdom \\ \and
Institute of Accounting and Administration, University of Aveiro, Portugal \\ \and
NUS High School of Math and Science, Singapore \\ \and
Qatar University, Doha, Qatar \\ \and
Information Technology University, Punjab, Pakistan \\ \and
Bristol Heart Institute, University of Bristol, Bristol, United Kingdom \\ 
$^*$Corresponding author: \email{Muhammad.bilal@uwe.ac.uk; Muhammad.bilal@bcu.ac.uk}}

\authorrunning{Bilal et al.}

\maketitle

\begin{abstract}
Coronary angiography analysis is a common clinical task performed by cardiologists to diagnose coronary artery disease (CAD) through an assessment of atherosclerotic plaque's accumulation. This study introduces an end-to-end machine learning solution developed as part of our solution for the MICCAI 2023 Automatic Region-based Coronary Artery Disease diagnostics using x-ray angiography imagEs (ARCADE) challenge, which aims to benchmark solutions for multivessel coronary artery segmentation and potential stenotic lesion localisation from X-ray coronary angiograms. We adopted a robust baseline model training strategy to progressively improve performance, comprising five successive stages of binary class pretraining, multivessel segmentation, fine-tuning using class frequency weighted dataloaders, fine-tuning using F1-based curriculum learning strategy (F1-CLS), and finally multi-target angiogram view classifier-based collective adaptation. 
Unlike many other medical imaging procedures, this task exhibits a notable degree of interobserver variability. 
Our ensemble model combines the outputs from six baseline models using the weighted ensembling approach, which our analysis shows is found to double the predictive accuracy of the proposed solution. The final prediction was further refined, targeting the correction of misclassified blobs. Our solution achieved a mean F1 score of $37.69\%$ for coronary artery segmentation, and $39.41\%$ for stenosis localisation, positioning our team in the 5th position on both leaderboards. This work demonstrates the potential of automated tools to aid CAD diagnosis, guide interventions, and improve the accuracy of stent injections in clinical settings.

\end{abstract}

\section{Introduction}
Coronary artery disease (CAD) is a major global health concern and a leading cause of death worldwide. The gold standard procedure for CAD diagnosis is invasive coronary angiography, which uses contrast materials and X-ray technology to visualise arterial lesions and assess blood flow to heart muscle in real-time. Such information plays a pivotal role in guiding intraventricular interventions, stent placements, and the planning of revascularisation procedures, relying on precise calculations of occlusions and affected artery segments (AS). However, X-ray coronary angiogram (XCA) like other imaging modalities faces a challenge in the form of interobserver variability, where angiographers often disagree on the severity of artery blockages\cite{zir1976interobserver}. To address this issue, there's a pressing need for digital tools to improve the reliability of XCA analysis. These tools hold the potential to significantly bolster the effectiveness of XCA for CAD detection and treatment strategies, addressing a critical gap in cardiac care.

\begin{figure}[!t]
    \centering
    \includegraphics[width=\textwidth]{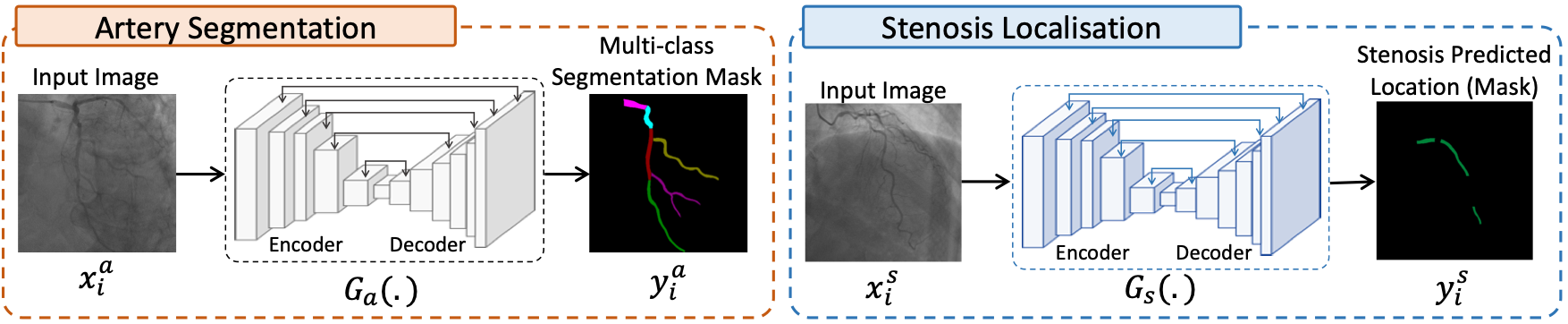}
    \caption{Our proposed method for segmentation of coronary arteries and stenotic lesions.}
    \label{fig:abs_fig}
\end{figure}

From an algorithmic standpoint, addressing the challenge of robust multivessel coronary artery segmentation and stenosis localisation in XCA images presents several complexities. These complexities arise due to factors such as patient-specific anatomical variations, the often unfavorable noise-to-signal ratio inherent to XCA images, and the presence of confounding background structures within the imaging scene. Various researchers have endeavored to tackle this issue. For instance, Gao et al. \cite{gao2022vessel} proposed an ensemble framework that combines gradient boosting decision trees and a deep forest classifier for binary segmentation of coronary arteries. In a similar vein, Danilov et al. \cite{danilov2021real} conducted a comparative study of modern neural network architectures, assessing their suitability for real-time stenosis localisation during procedures. However, a significant portion of existing research is hindered by limitations, including small datasets, impractical binary segmentation methods for clinical settings, and insufficient support for clinicians in their routine XCA analysis tasks.

The MICCAI 2023 Automatic Region-based Coronary Artery Disease diagnostics using x-ray angiography imagEs (ARCADE) challenge enabled our research endeavour by providing the essential framework for benchmarking our solutions in coronary artery segmentation and stenosis localisation. This competition has attracted biomedical AI researchers from around the world, all striving to develop cutting-edge solutions for this problem. To facilitate the participants' efforts, the challenge organisers have generously provided a comprehensive dataset consisting of 1,500 meticulously annotated images of coronary AS and an additional 1,500 images segmenting the locations of stenotic plaques. The labels for both multivessel instance segmentation and stenosis localisation tasks adhere to the SYNTAX Score\footnote{\url{https://syntaxscore.org/index.php/tutorial/definitions/14-appendix-i-segment-definitions}} definitions, wherein each region of the vessel tree is assigned a name and an ordinal number based on its specific location. This dataset is divided into three subsets: a \textit{training set} containing 1,000 images, a \textit{phase 1 validation set} comprising 200 images, and a \textit{final phase validation set} with 300 images. The evaluation of a participant's model is carried out using the F1 score, calculated for each individual image. The overall F1 score for a team is then determined as the average F1 score across all images. When multiple teams achieve the same F1 score, inference time serves as the tiebreaker.  

Our proposed solution comprises a multifaceted approach aimed at enhancing the accuracy of coronary angiogram analysis. We employ image transformations specifically tailored to mitigate noise in XCA images. Our solution leverages six baseline models selected from state-of-the-art (SOTA) deep learning (DL) architectures. The training unfolds in five stages: binary class segmentation pretraining, multilabel vessel segmentation training, two subsequent fine-tuning stages using class frequency dataloaders (to learn minority classes), and F1-based curriculum learning (to address difficult classes). Finally, collaborative classifier-based fine-tuning accurately predicts AS based on the given angiogram view. Weighted ensembling significantly improved the output quality, followed by morphology operations for small segment removal and gap filling, further refining the results. 

Our solution exhibits novelty in several areas. Firstly, we utilise inter-class difficulty-aware stratified sampling, enabling our baseline models to maximise the utility of the limited dataset. Secondly, our systematic baseline model training process incorporates various modelling strategies in a meaningful order, including a unique collaborative multi-target training stage. This facilitates the learning of view-specific AS by considering angiogram acquisition angles. This training process worked on all baseline models and progressively improved their accuracy, regardless of their underlying models' architectures.

\section{Related Work}
\label{sec:related}
Deep Ensemble Learning (DEL) has emerged as a potent technique in medical image analysis, notably excelling in multivessel coronary artery segmentation, disease localisation, and its broader applications in radiology and pathology. DEL's strength lies in aggregating multiple weak models to improve segmentation precision and accuracy. This section provides a concise review of pertinent literature, serving to contextualise our research within the existing landscape of image analysis. By summarising key findings and contributions from related papers, we position our work as a progressive addition to the ongoing dialogue in this crucial domain.

Nobre et al. \cite{nobre2023coronary} proposed an AI solution to segment XCA images from four medical centres encompassing patients who underwent CAG, PCI, or invasive assessments. Their approach showcased a notable increase in multiple segmentation metrics, including overlap accuracy, sensitivity, and Dice Score. Furthermore, the Global Segmentation Score (GSS) exhibited alignment with previous results, thus serving as a validation of AI segmentation effectiveness in the realm of XCA analytics. The authors highlight critical clinical applications concerning coronary artery revascularisation. Tao et al. \cite{tao2022lightweight} proposed a Bottleneck Residual U-Net (BRU-Net), a lightweight model for XCA segmentation. Differing from the traditional U-Net, BRU-Net incorporates bottleneck residual blocks to enhance computational efficiency. The CLAHE pre-processing technique is also used to improve performance. Nevertheless, BRU-Net occasionally mislabels background as vessels and vice versa, impacting accuracy. Data annotation challenges result in unmarked thin vessels being segmented, causing discrepancies. The authors emphasise the need for larger, improved coronary angiography datasets to bolster coronary artery disease diagnosis.

Additionally, Gao et al. \cite{gao2022vessel} proposed an innovative method for delineating coronary blood vessels by leveraging a combination of DL and filter-based features, all integrated into an ensemble framework that utilises Gradient Boosting Decision Tree (GBDT) and Deep Forest classifiers. Compared to traditional deep neural networks, this ensemble approach demonstrates superior performance across various metrics, including precision, sensitivity, specificity, F1 score, AUROC (Area Under the Receiver Operating Characteristic curve), and IoU (Intersection over Union) scores. Tmenova et al. \cite{tmenova2019cyclegan} proposed enhancing the realism of vascular images generated from a cardiorespiratory simulator. They employed CycleGAN to mimic real angiograms. The evaluation focused on the consistency and vessel preservation of the enhanced images, resulting in an average structural similarity (SSIM) score of 0.948. These findings suggest that CycleGAN serves as a potent tool for synthetic XCA data generation. This approach is particularly relevant due to the need to replicate the complex physiology and patterns observed in X-ray angiography, as well as the scarcity of data mentioned before.

\section{Proposed Method}
\label{sec:method}
\begin{table}[!h]
\label{tab:data_stats_seg}
\centering
\caption{Summary of class-wise statistics of samples in Arcade dataset.}
\resizebox{0.85\textwidth}{!}{%
\begin{tabular}{@{}ccrrrrrr@{}}
\toprule
\begin{tabular}[c]{@{}c@{}}Class \\ ID\end{tabular} &
  \begin{tabular}[c]{@{}c@{}}Class \\ Name\end{tabular} &
  \begin{tabular}[c]{@{}r@{}}Total Segments\\ Count\end{tabular} &
  \begin{tabular}[c]{@{}r@{}}Total Segment\\ Pixels\end{tabular} &
  \begin{tabular}[c]{@{}r@{}}Min Segment \\ Size (Pixels)\end{tabular} &
  \begin{tabular}[c]{@{}r@{}}Max Segment \\ Size (Pixels)\end{tabular} &
  \begin{tabular}[c]{@{}r@{}}Avg Segment \\ Size (Pixels)\end{tabular} &
  \begin{tabular}[c]{@{}r@{}}Dataset Size \\ (\% Pixels)\end{tabular} \\ \midrule \midrule
\multicolumn{8}{c}{Data description: artery segmentation dataset} \\ \hline 
0  & Background & 1,000 & 261,957,856 & 9   & 262,144 & 248,536.87 & 97.0839 \\
1  & 1          & 404  & 650,624    & 4   & 4,549   & 1,610.46   & 0.2411  \\
2  & 2          & 386  & 666,498    & 79  & 4,986   & 1,726.68   & 0.247   \\
3  & 3          & 394  & 653,260    & 9   & 5,168   & 1,658.02   & 0.2421  \\
4  & 4          & 346  & 478,564    & 8   & 5,394   & 1,383.13   & 0.1774  \\
5  & 5          & 529  & 574,036    & 52  & 2,827   & 1,085.13   & 0.2127  \\
6  & 6          & 498  & 705,086    & 64  & 3,123   & 1,415.84   & 0.2613  \\
7  & 7          & 348  & 600,557    & 5   & 4,035   & 1,725.74   & 0.2226  \\
8  & 8          & 323  & 490,696    & 16  & 3,332   & 1,519.18   & 0.1819  \\
9  & 9          & 189  & 408,668    & 11  & 7,546   & 2,162.26   & 0.1515  \\
10 & 9a         & 85   & 136,233    & 339 & 4,218   & 1,602.74   & 0.0505  \\
11 & 10         & 31   & 23,793     & 13  & 1,856   & 767.52    & 0.0088  \\
12 & 10a        & 1    & 905       & 905 & 905    & 905.0     & 0.0003  \\
13 & 11         & 310  & 443,786    & 6   & 3,202   & 1,431.57   & 0.1645  \\
14 & 12         & 52   & 114,480    & 501 & 6,332   & 2,201.54   & 0.0424  \\
15 & 12a        & 138  & 305,307    & 288 & 6,039   & 2,212.37   & 0.1131  \\
16 & 13         & 294  & 462,611    & 53  & 4,620   & 1,573.51   & 0.1714  \\
17 & 14         & 106  & 246,130    & 12  & 7,379   & 2,321.98   & 0.0912  \\
18 & 14a        & 45   & 78,563     & 636 & 4,996   & 1,745.84   & 0.0291  \\
19 & 15         & 31   & 44,301     & 469 & 2,541   & 1,429.06   & 0.0164  \\
20 & 16         & 264  & 340,256    & 229 & 4,772   & 1,288.85   & 0.1261  \\
21 & 16a        & 54   & 59,062     & 179 & 2,502   & 1,093.74   & 0.0219  \\
22 & 16b        & 61   & 77,381     & 92  & 2,646   & 1,268.54   & 0.0287  \\
23 & 16c        & 31   & 33,594     & 176 & 2,063   & 1,083.68   & 0.0125  \\
24 & 12b        & 57   & 126,006    & 236 & 6,248   & 2,210.63   & 0.0467  \\
25 & 14b        & 149  & 148,024    & 4   & 3,269   & 993.45    & 0.0549  \\ \midrule \midrule

\multicolumn{8}{c}{Data description: stenosis localisation dataset} \\ \hline

0 &
  Background &
  1,000 &
  262,144,000 &
  262,144 &
  262,144 &
  262,144.0 &
  99.2783 \\
26 &
  Stenosis &
  1,550 &
  1,905,633 &
  96 &
  7,752 &
  1,229.44 &
  0.7217 \\ \bottomrule 
\end{tabular}}
\end{table}


\subsection{Data Description}
The ARCADE challenge comprises two distinct datasets, each consisting of 1,500 XCA images, purposefully curated for the intricate task of coronary artery segmentation and stenosis localisation. The training set included 1,000 images whereas two sets of validation XCA images were held out by the competition organisers, including the phase 1 validation set involving 200 images, and the final phase set comprised an additional 300 images. To gain insights into the dataset and to make informed modelling decisions, we conducted exploratory mask analysis. Table \ref{tab:data_stats_seg} provides key summary statistics for Arcade datasets. We identified a total of 6,180 coronary artery segmentation masks across 25 segmentation classes, adhering to the SYNTAX Score definition. However, it's important to note that the background pixel (class 0) accounts for a staggering 97.08\% of the entire dataset, leaving less than 3\% of pixels available to inform the learning of visual features for multivessel segmentation. Among the non-background classes, 13 classes (6, 2, 1, 3, 7, 5, 4, 8, 13, 11, 9, 16, 12a) contained over 100 segmentation masks, while 12 classes (14, 9a, 12b, 14b, 12, 14a, 16b, 15, 16a, 10, 16c, 10a) had fewer than 100 masks in the training data. Notably, some segments, such as class 10a, were represented by only a single image in the dataset. This highlights a significant class imbalance, primarily within the background class (at pixel level) but also among other AS, with varying ratios (at sample level). This class imbalance introduces several challenges, including a propensity for biased learning and difficulty in learning minority classes.

\begin{figure}[!t]
    \centering
    \includegraphics[width=0.95\linewidth]{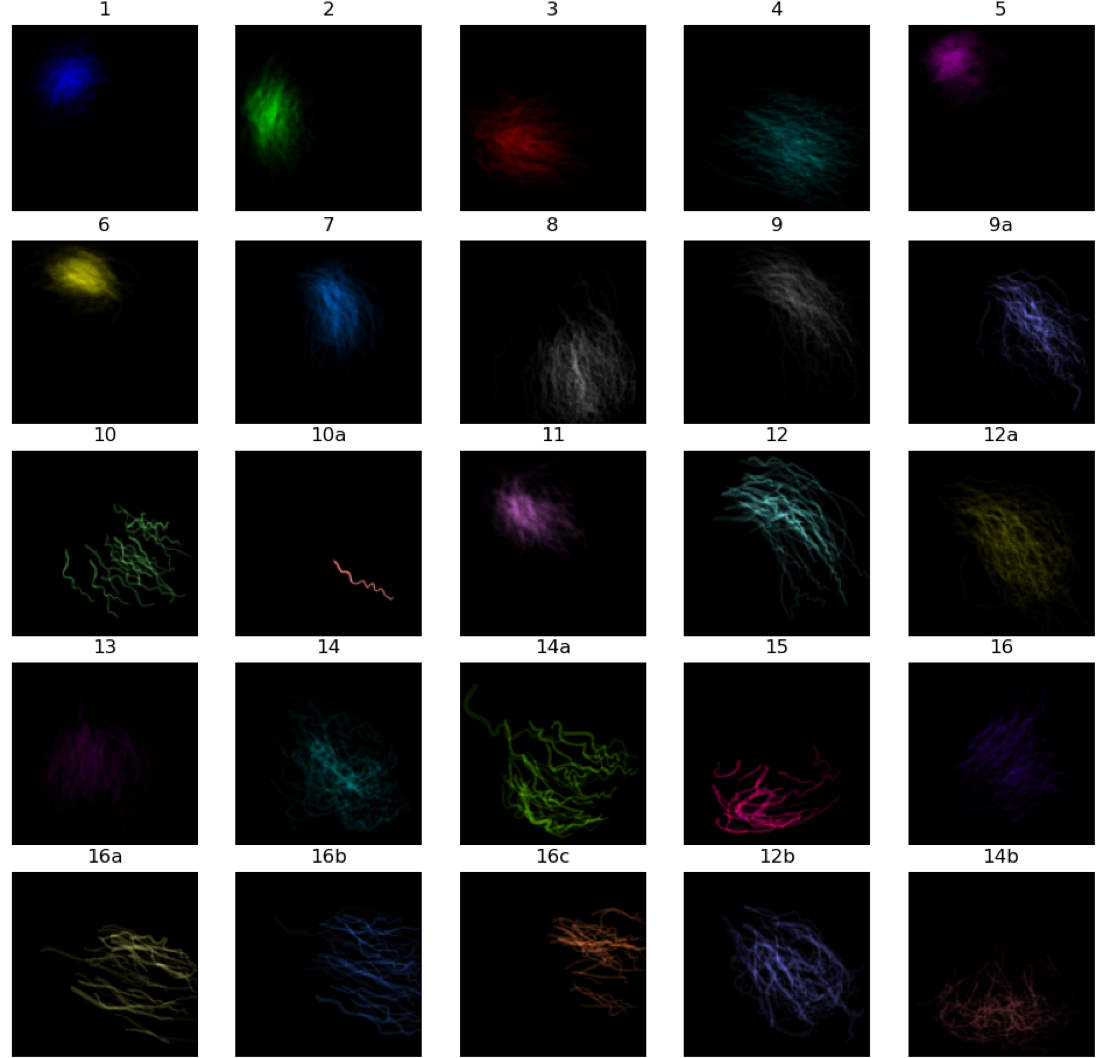}
    \caption{Illustration of different artery segments (AS) with mean variations in the dataset. \textit{$^*$Colors are used for illustrative purposes.}}
    \label{fig:data_exp}
\end{figure}

Furthermore, the dataset exhibits intraclass segment variability, with segment sizes within the same class varying substantially, ranging from single pixels to several thousand pixels forming a segmentation mask. In Fig. \ref{fig:data_exp} we highlight this diversity by calculating the mean of each artery segment in the dataset to get an average of their representation across the dataset. This variation in segment sizes poses inherent challenges, as classes with significant segment size disparities may be difficult to learn and also require intelligent mechanisms to ensure similar difficulty is reflected by the training and validation sets of 1,000 images. An important limitation of this dataset is the issue of overlapping segments, particularly at the borders where two segments interleave. While the original annotations were provided in JSON files to allow multiple classes to represent a single pixel location in a mask, this information was lost when we serialised and saved masks as PNG files. Perhaps the most formidable challenge presented by the dataset lies in the fact that not all coronary arteries visible in the XCA view are segmented. Instead, only a small set of arteries that were the focus at the given time in XCA procedure are segmented. This characteristic renders the modelling task exceptionally challenging, as it necessitates the ability to discern and segment relevant arteries amidst a complex and cluttered background.

\subsection{Proposed Solution}
\begin{figure}[!t]
    \centering
    \includegraphics[width=\textwidth]{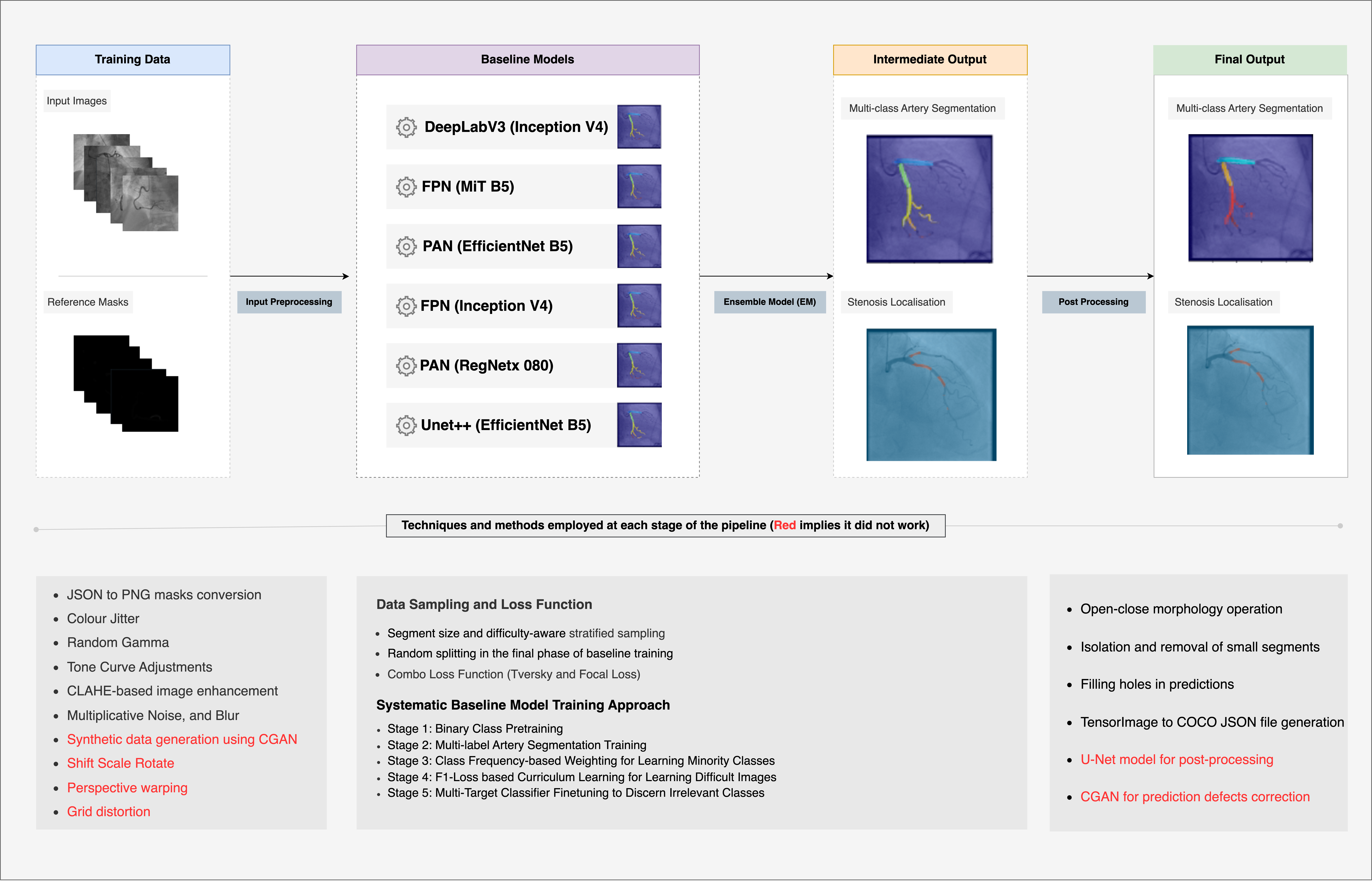}
    \caption{Our proposed coronary artery and stenosis segmentation methodology}
    \label{fig:method}
\end{figure}

Our proposed methodology for joint segmentation of artery segmentation and stenosis localisation consists of four major steps that include: (1) Input Preprocessing; (2) Baseline Model Development; (3) Ensemble Model Development; and (4) Post processing. The pipeline of our proposed solution is shown in Fig. \ref{fig:method} and described next. 

\noindent \textbf{Input Preprocessing:} Data preprocessing plays a crucial role in the ML pipeline, addressing issues within the input XCA data. Our preprocessing steps included converting JSON-formatted COCO channel-encoded masks into 2D class-encoded masks, although some information was lost due to overlaps of artery segments around their boundaries. For binary class pretraining of the multivessel segmentation stage, as well as stenosis localisation model training, we transformed mask contents into binary images, designating $0$ for the background and $255$ for the foreground by replacing pixels greater than $0$ with $255$. We further improved input image quality by applying various techniques, such as contrast-limited histogram equalisation (CLAHE), Gabor filtering, random gamma adjustments, and tone curve enhancements. To match the difficulty of the validation set, we also introduced multiplicative noise and a slight blur.

\noindent \textbf{Baseline Model(s) Development:} In this section, we describe the development of our baseline models. We considered training SOTA segmentation models, including DeepLabV3+ \cite{chen2018encoder}, FPN \cite{kirillov2017unified}, PAN \cite{li2018pyramid}, and UNet++ \cite{zhou2018unet++}. These models are integrated with different encoders, encompassing DeepLabV3+ with InceptionV4, FPN with MiT B5, PAN with EfficientNet-B5, FPN with InceptionV4, PAN with RegNetx080, and UNet++ with an EfficientNetB5 backbone. Our approach to training baseline models adheres to a systematic strategy, which we will elaborate on further in this section. Additionally, we will elucidate the key design choices that played a pivotal role in our model development process, including a novel data-splitting strategy designed to create a challenging validation set and the implementation of a combo loss function aimed at facilitating the learning of robust imaging features from a dataset of such complexity.

\vspace{1mm}
\noindent \textit{Intraclass Difficulty-aware Stratified Sampling:}
We employed a novel sampling strategy to balance class distribution in both the training and validation sets. Let $C$ denote the total number of classes, $N$ represent the total number of samples, $S_i$ denote the size of segment $i$ (where $i$ ranges from 1 to $N$), and $n_i$ represent the number of samples belonging to class $i$. We partitioned the dataset into training and validation sets, ensuring class balance while considering segment size. To allocate $V$ samples to the validation set, we considered the desired number of segments to assign to the validation set, defined as $S_i < S_t$, where $S_t$ represents a specified threshold value. We aimed to distribute smaller segments to the validation set while maintaining class balance, using class proportions $P_i$, calculated as $P_i = \frac{N}{n_i}$. We followed these steps: (1) Calculated the number of segments ($V_i$) with $S_i < S_t$ for each class; (2) Assigned $V_i$ segments from each class $i$ to the validation set; and (3) Utilised class proportions $P_i$ to determine how many segments $V_i$ from each class should be assigned to the validation set, ensuring class equilibrium. We compute $V_i$ as follows:

\begin{equation}
    V_i = \frac{{P_i \times V}}{{\sum_{j=1}^{C} P_j}}
\end{equation}

Here, ${\sum_{j=1}^{C} P_j}$ represents the sum of class proportions for all classes. This approach balances class distribution and effectively allocates smaller segments to the validation set, guiding the model's learning process to accommodate varying difficulty levels, ultimately enhancing experimental effectiveness.

\vspace{1mm}
\noindent \textit{Combo Loss Function:} We combined Focal Loss and Tversky losses to address distinct challenges in the ARCADE dataset. Focal Loss is chosen for its effectiveness in handling imbalanced datasets, specifically to enhance the baseline models' performance on minority classes. Likewise, Tversky Loss was introduced to tackle difficult classes with significant segment size variability. We found a remarkable synergy between these loss components by weighting both losses with a balance parameter, represented as $\alpha$ (set to 0.5), achieving a harmonious fusion of their respective strengths. We explored several gamma ($\gamma$) values (ranging from 2 to 4) in both loss components, adapting the model's focus to harder examples as training progressed. Let $\mathcal{L}_{F}$ denote the Focal Loss, $\mathcal{L}_{T}$ represents the Tversky Loss, and $\alpha$ be the balance parameter. The combo loss of function $\mathcal{L}_{C}$ can be mathematically expressed as $\mathcal{L}_{C} = \alpha \times \mathcal{L}_{F} + (1 - \alpha) \times \mathcal{L}_{T}$. 
Where, $\alpha$ regulates the weightage assigned to each loss, allowing for a dynamic adjustment of models to focus on class imbalance and learning challenging segments. This flexible combination not only harmonises the strengths of these loss functions but also enables adaptability during training by considering various values of $\alpha$ and $\gamma$.


\noindent \textit{Systematic Baseline Model Training Approach:} Our solution relies on baseline models for coronary artery segmentation and stenosis localization. We selected six top-performing architectures and encoder combinations from a list of pretrained models, as described above. The subsequent subsections provide a detailed explanation of each of these stages.

\vspace{1mm}
\noindent \textit{Stage 1: Binary Class Pretraining:} We initiated the training of our baseline models through transfer learning. Notably, our initial experiments revealed the inadequacy of using pretrained IMAGENET models. This limitation stemmed from the fundamental differences between the objects in the IMAGENET dataset and the anatomical structures commonly found in coronary artery angiograms. Subsequently, we trained a binary class segmentation model to learn basic spatial patterns of coronary arteries, considering them as the foreground class and non-arterial pixels as the background class. This domain-specific binary class pretraining offered several advantages. By simplifying the initial segmentation task, the baseline models converged quickly during training. Additionally, this pretraining endowed the subsequent multivessel segmentation models with fundamental spatial awareness of the coronary artery tree. As a result, the baseline models significantly improved their performance on segmentation tasks, even after just the first epoch of training. The mean F1 score of the binary class segmentation models after $80$ epochs of training was $85.34\%$.

\vspace{1mm}
\noindent \textit{Stage 2: Multivessel Artery Segmentation Training:} This stage marks the beginning of training our baseline multi-vessel segmentation models. We initiated it using binary class pretrained models as a starting point. Notably, the baseline models quickly began to discern coronary artery segments, resulting in a performance boost of 10 points compared to pretraining with IMAGENET models. On the training set, we achieved an F1 score exceeding $50\%$, a notable improvement from the previous performance plateau of around $40\%$. The training of the multivessel segmentation models spanned $80$ epochs, with an initial learning rate of $1e-3$ for pretraining, followed by $50$ epochs of fine-tuning. During the fine-tuning phase, we unfroze the entire model architecture and applied much smaller learning rates, ranging from $1e-3/400$ to $1e-3/4$. The mean F1 score of the baseline models consistently hovered around $15\%$ across both evaluation phases.

\vspace{1mm}
\noindent \textit{Stage 3: Class Frequency-based Weighting for Learning Minority Classes:}
Our initial evaluation revealed that baseline models struggled to predict minority classes, as evidenced by their lower F1 scores. To address this challenge, we introduced a novel scoring mechanism designed to assign higher probabilities to rare class XCA images using weighted dataloaders, thereby enhancing their representation in mini-batches creation during training. 

We calculate the class frequency, denoted as \(F_c\), for each class, which is defined as the proportion of segments belonging to a specific class within the entire dataset: $F_{c} = \frac{|j|}{|\mathcal{D}|}$, where $|j|$ is the number of segments in class $j$, and $|\mathcal{D}|$ is the total number of class segments in the dataset. To convert these frequencies into scores for each class, we take the square root of the class frequency, defined as $S_c = \sqrt{F_c}$.

To compute a single score for an XCA image $(x)$, since most samples had several class segments, we computed the reciprocals of \(S_c\) values for all unique class segments present in the XCA mask, denoted as \(U_{c}(x)\), and selected the lowest class score as the weight for the XCA image to guide mini-batch sampling, denoted as \(S_{x}\):

\begin{equation}
S_{x} = \min_{x \in U_{c}(x)}\left(\frac{1}{S_{c}}\right)
\end{equation}

This scoring approach reduces the probability of selecting frequent classes, allowing dataloaders to encounter minority classes more frequently and enabling the models to learn to distinguish rare segments. While this resulted in a slight reduction in the overall performance, the baseline models exhibit improved F1 scores for minority classes.

\noindent \textit{Stage 4: F1-Loss based Curriculum Learning for Learning Difficult Images:}
In this stage, we introduce a novel approach called "F1-loss-based Curriculum Learning Strategy (F1-CLS)" to further enhance the learning capabilities of our models to address the difficulty induced by images with significant segment size variability. By utilising the F1 score as a loss function, our aim is to employ self-directed curriculum learning to guide the model's learning process toward these challenging samples. We define the F1 score (\(F_i\)) for each XCA image (\(F_i(x)\)) as $F_i = \frac{2 \times P_i \times R_i}{P_i + R_i}$, where \(P_i\) represents the precision of the model's predictions for \(F_i(x)\), and \(R_i\) represents the recall of the model's predictions for \(F_i(x)\). The precision and recall values are computed based on the true positive (TP), false positive (FP), and false negative (FN) predictions for each class in \(F_i(x)\).

We implemented curriculum learning to gradually introduce difficult images into the training process. The training loop starts with a subset of the dataset that contains images with less segment size variability. As the model's performance improves, we progressively increase the frequency of complex images in mini-batches, specifically those on which the model performs poorly at the beginning to initiate learning from them. This process allows the model to learn from easier samples initially and gradually tackle the more challenging ones, resulting in a more robust and effective learning process. F1-CLS has proven to be instrumental in improving the model's performance, particularly on difficult classes with varying segment sizes. Through this approach, we enable our models to excel in capturing the intricacies of coronary artery segmentation, even in the presence of challenging samples, leading the models to achieve a mean F1 score of around 18\% on both validation phases.

\vspace{1mm}
\noindent \textit{Stage 5: Multi-Target Classifier Finetuning for Plane-Based Segmentation:}
In the preceding training stages, our baseline models improved their segmentation accuracy on minority class samples and difficult images. However, a new challenge emerged: these models sometimes predicted arteries where they should not be present in certain angiogram planes. This stage introduces an innovative approach to incorporate angiogram plane knowledge using multi-target training to enhance the performance of our baseline models.

Our research identified 11 acquisition planes for XCA images, each associated with specific classes of the coronary artery tree. Collaborating closely with our clinical partners, we annotated ARCADE training images with their respective acquisition planes. Subsequently, we developed a classifier capable of predicting the acquisition plane for a given image with $75\%$ accuracy. We then used this classifier to further refine our baseline segmentation models through a multi-target model. This model leverages both the view classifier and the baseline segmentation model, training them further using a multi-target loss function that combines cross-entropy and our proposed combo loss $\mathcal{L}_{C}$ discussed earlier.

This collaborative training significantly boosted the performance of both models. The view classifier achieved an impressive accuracy score of $84\%$, while the performane of baseline models improved by an average of three points. Our evaluation revealed that these models now predict the correct combinations of artery segments for the specific angiogram plane, effectively eliminating irrelevant predictions. This final stage of training marks a crucial milestone in our journey toward achieving accurate and clinically relevant coronary artery segmentation.


\noindent \textbf{Ensemble Model Development:} In our quest to maximise the predictive performance of our solution, we delved into ensemble learning as a potent strategy. Ensemble learning involves integrating multiple weak models to create a single, potentially highly robust model. We harnessed the capabilities of six diverse baseline models, each trained with distinct deep learning architectures and encoders, with the aim of leveraging the inherent variance in their outputs. While our baseline models produced only modest predictions, the ensemble approach astutely harnessed the complementary strengths of individual models, culminating in a potent model. The predicted outputs ($O_1, \ldots, O_n$), one from each baseline model, correspond to their predicted masks on the XCA image. To assign appropriate importance to each model's prediction, we introduced a weight vector based on the models' performance on the training dataset to quantify their influence within the ensemble. Mathematically, our ensemble output is computed as follows:

\begin{equation}
\hat{f}_{ens}(x) = \frac{1}{B} \sum_{i=1}^{B} \hat{f}_i(x)
\end{equation}

Here, $\hat{f}_{ens}(x)$ represents the ensembled output, achieved by averaging the individual mask predictions of baseline models ($\hat{f}_1(x), \ldots, \hat{f}_M(x)$). The variable $M$ signifies the number of baseline models in our ensemble.

Ensemble learning emerged as a highly effective strategy for enhancing predictive performance. It empowered our solution to exploit diverse DL architectures and encoders, resulting in a remarkable improvement of over 50\% compared to individual models. To determine the optimal combination of models within the ensemble, we devised an algorithm for the systematic exploration of various model combinations. By objectively selecting the best-performing combinations, we further refined our ensemble, ultimately achieving superior results.

\begin{figure}[!t]
    \centering
    \includegraphics[width=0.9\linewidth]{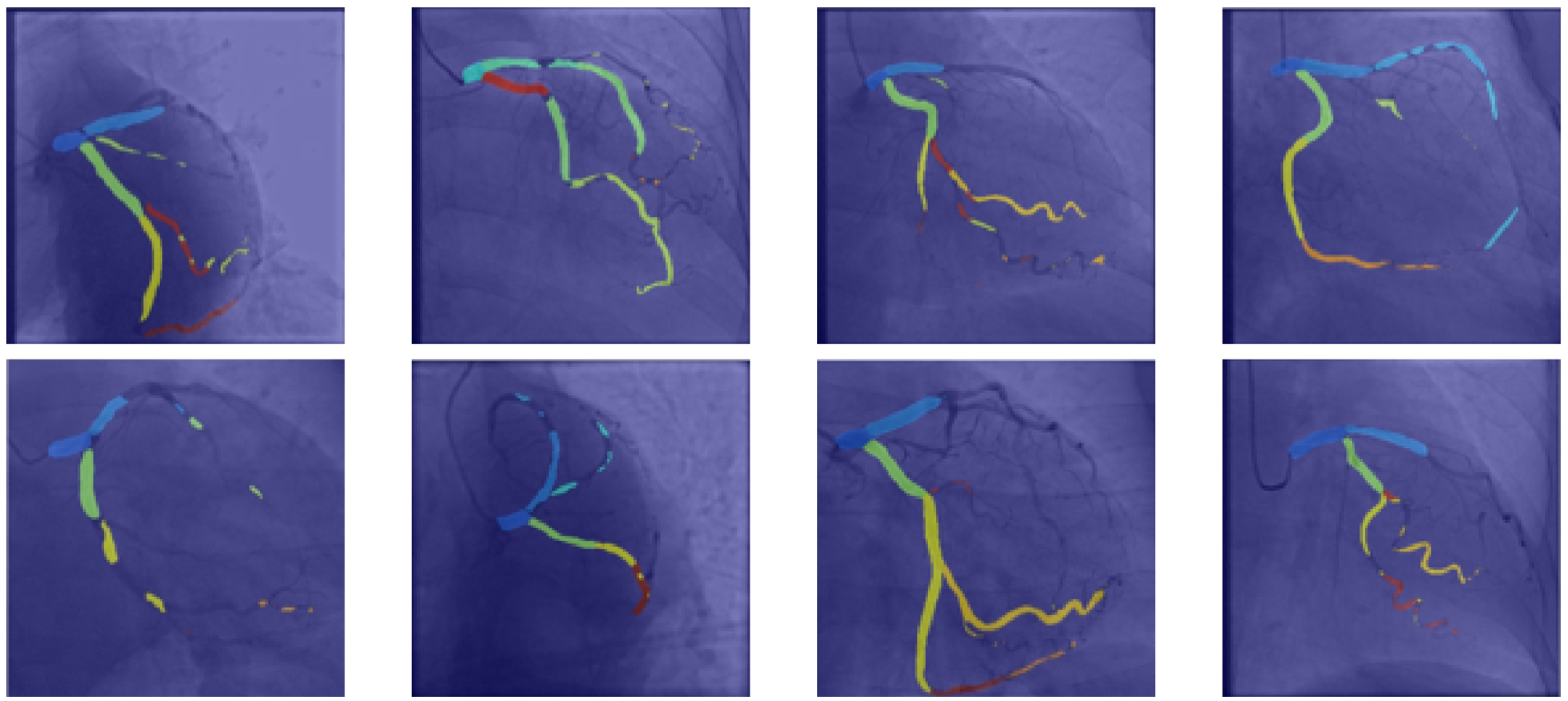}
    \caption{Examples of errors in predicted masks (including unrealistic segment sizes, sporadic voids, and amalgamated segments).}
    \label{fig:pred_defects}
\end{figure}

\noindent \textbf{Post Processing for Error Correction:}
In the end, we focused on rectifying errors within the predicted segmentation masks to enhance the overall output quality. As depicted in Fig. \ref{fig:pred_defects}, our ensemble predictions exhibited various errors, appearing in diverse forms such as noisy pixels along borders or at a distance from the primary AS, segments with unrealistic sizes (either excessively large or small), gaps within AS, sporadic voids within segments, and instances where one artery segment encompassed another. To address these challenges, we devised a comprehensive post-processing pipeline. This pipeline included a two-step open-close morphology operation, followed by the identification and removal of unrealistic AS. Often underestimated, post-processing played a pivotal role in enhancing the quality of our segmentation masks. As a result, our solution achieved a noteworthy improvement of $3.74\%$ and $3.20\%$ in performance, both on the phase 1 and final phase validation sets, over the ensemble outputs.

\begin{table}[!t]
\centering
\caption{Comparative analysis of different training strategies and their impact on the baseline models and ensemble model for AS segmentation in terms of mean F1 score using phase 1 and 2 validation sets. }

\label{tab:results}
\resizebox{\textwidth}{!}{%
\begin{tabular}{clrrrrrrrr}
  &                              &         &         &         &         &         &         &         &         \\ \hline
\multicolumn{1}{c|}{\begin{tabular}[c]{@{}c@{}}Phase \\ Id\end{tabular}} &
  \multicolumn{1}{l|}{\begin{tabular}[c]{@{}l@{}}Phase\\ Name\end{tabular}} &
  \multicolumn{1}{r|}{\begin{tabular}[c]{@{}r@{}}FPN \\ Mix ViT \\ B5\end{tabular}} &
  \multicolumn{1}{r|}{\begin{tabular}[c]{@{}r@{}}PAN \\ EfficientNet \\ B5\end{tabular}} &
  \multicolumn{1}{r|}{\begin{tabular}[c]{@{}r@{}}FPN \\ Inception \\ V4\end{tabular}} &
  \multicolumn{1}{r|}{\begin{tabular}[c]{@{}r@{}}DeepLabV3+ \\ EfficientNet \\ B5\end{tabular}} &
  \multicolumn{1}{r|}{\begin{tabular}[c]{@{}r@{}}PAN \\ RegNetX \\ 080\end{tabular}} &
  \multicolumn{1}{r|}{\begin{tabular}[c]{@{}r@{}}Unet++ \\ EfficientNet \\ B5\end{tabular}} &
  \multicolumn{1}{r|}{Ensemble} &
  \begin{tabular}[c]{@{}r@{}}Post \\ Processing\end{tabular} 
  \\ \midrule \midrule \multicolumn{10}{c}{Phase 1 Evaluation using Mean F1 Score} \\ \hline 
1 & Binary Class Pretraining        & 6.74\%  & 5.46\%  & 8.91\%  & 6.38\%  & 5.53\%  & 3.09\%  & 11.69\% & 13.20\% \\
2 & Multvessel Segmentation      & 19.85\% & 13.75\% & 26.26\% & 13.17\% & 17.51\% & 9.36\%  & 33.97\% & 37.87\% \\
3 & Class Frequency Dataloaders  & 19.58\% & 12.94\% & 22.23\% & 16.58\% & 16.42\% & 12.47\% & 35.55\% & 39.52\% \\
4 & F1-based Curriculum Learning & 21.97\% & 16.71\% & 22.39\% & 21.40\% & 18.03\% & 15.82\% & 36.53\% & 39.74\% \\
5 & Multitarget Finetuning       & 21.97\% & 18.32\% & 20.46\% & 20.51\% & 17.52\% & 16.65\% & 35.52\% & 39.39\%
  \\ \midrule \midrule \multicolumn{10}{c}{Final Phase Evaluation using Mean F1 Score} \\ \hline 
1 & Binary class pretraining     & 6.13\%  & 6.49\%  & 8.34\%  & 7.33\%  & 5.98\%  & 4.17\%  & 10.33\% & 12.91\% \\
2 & Multvessel Segmentation      & 25.41\% & 19.26\% & 25.04\% & 13.94\% & 15.62\% & 12.45\% & 32.22\% & 35.70\% \\
3 & Class Frequency Dataloaders & 24.59\% & 20.20\% & 22.27\% & 19.40\% & 16.78\% & 14.74\% & 33.10\% & 36.50\% \\
4 & F1-based Curriculum Learning & 24.68\% & 21.46\% & 21.45\% & 21.43\% & 17.86\% & 18.42\% & 33.37\% & 36.51\% \\
5 & Multitarget Finetuning       & 24.84\% & 22.04\% & 24.88\% & 22.20\% & 17.85\% & 18.42\% & 34.89\% & 37.65\% \\ \hline
\end{tabular}%
}
\end{table}

\section{Results and Discussion}
\label{sec:results}

\subsection{Main Results (using Phase 1 and 2 Validation Datasets)}
In this section, we present the results of our proposed solution for training coronary artery segmentation and stenosis localisation models. Table \ref{tab:results} present our results on both validation set (i.e., phase 1 and phase 2 provided by Arcade challenge). From the table, it is evident that our proposed systematic training methodology produced improved models, which were subsequently enhanced by ensemble techniques and further refined through post-processing in both evaluation phases. This highlights the effectiveness of our meticulously designed multi-staged training approach was found to systematically enhance the performance of models across different architectural families. Below, we delve into the results obtained in each stage of our systematic strategy, emphasising the stark contrast between our ensemble model and the baseline models. Additionally, we highlight the efficacy of the proposed post-processing tasks in improving the ensemble output.

\subsection{Effect of Systematic Training Strategy}

In \textit{Stage 1}, \textit{the binary class segmentation pretraining}---the performance of baseline models was as expected, with the ability to correctly identify approximately 10\% of artery pixels without training. Even the ensemble of binary class models achieved a modest F1 score of 11.69\%. Subsequent post-processing techniques were applied, resulting in a noticeable improvement, with the mean F1 score reaching 13.20\%. These results underscore the inherent difficulty of the task, as individual models struggled to achieve satisfactory accuracy.

In \textit{Stage 2}, we introduced multilabel adaptation of the binary class model, some of these models, such as FPN Mix ViT B5, started to learn artery pixels, surpassing a mean F1 score of 19\%. The ensemble models demonstrated a remarkable leap, achieving an F1 score of 33.97\%. Post-processing further enhanced the score to 37.87\%. This considerable increase in accuracy highlighted the limitations of individual models in comprehending the intricate coronary artery structures.  

In \textit{Stage 3}, we introduced class frequency fine-tuning, emphasising the importance of accurately segmenting minor classes. While baseline models' accuracy slightly decreased compared to the previous stage, the ensemble model exhibited superior performance, achieving an F1 score of 35.55\%, which increased to 39.52\% with post-processing. This stage underscored the value of addressing minor classes, a task where baseline models often faltered.

\textit{Stage 4} introduced F1-CLS, focusing on challenging classes. It marked a turning point, as the ensemble model's F1 score surged to 36.53\%. Post-processing further refined it to 39.74\%. In contrast, the baseline models continued to struggle with challenging classes, reaffirming the limitations of individual models.

Finally, in \textit{Stage 5}, multi-target fine-tuning was employed to refine the segmentation model using angiography view predictions. The ensemble model achieved an F1 score of 35.52\%, with post-processing increasing it to 39.39\%. In contrast, the baseline models could not adapt to the clinical relevance aspect of the task, further highlighting their limitations. Note the results discussed above are based on phase 1 validation data (as presented in the top half of Table \ref{tab:results}). However, we observed a slight drop in the overall performance of different segmentation models and the techniques employed, when models were evaluated using phase 2 validation data (see bottom half of Table \ref{tab:results}).

\subsection{Effect of Ensemble Learning and Post Processing}

\begin{figure}[!t]
    \centering
    \includegraphics[width=1.0\textwidth]{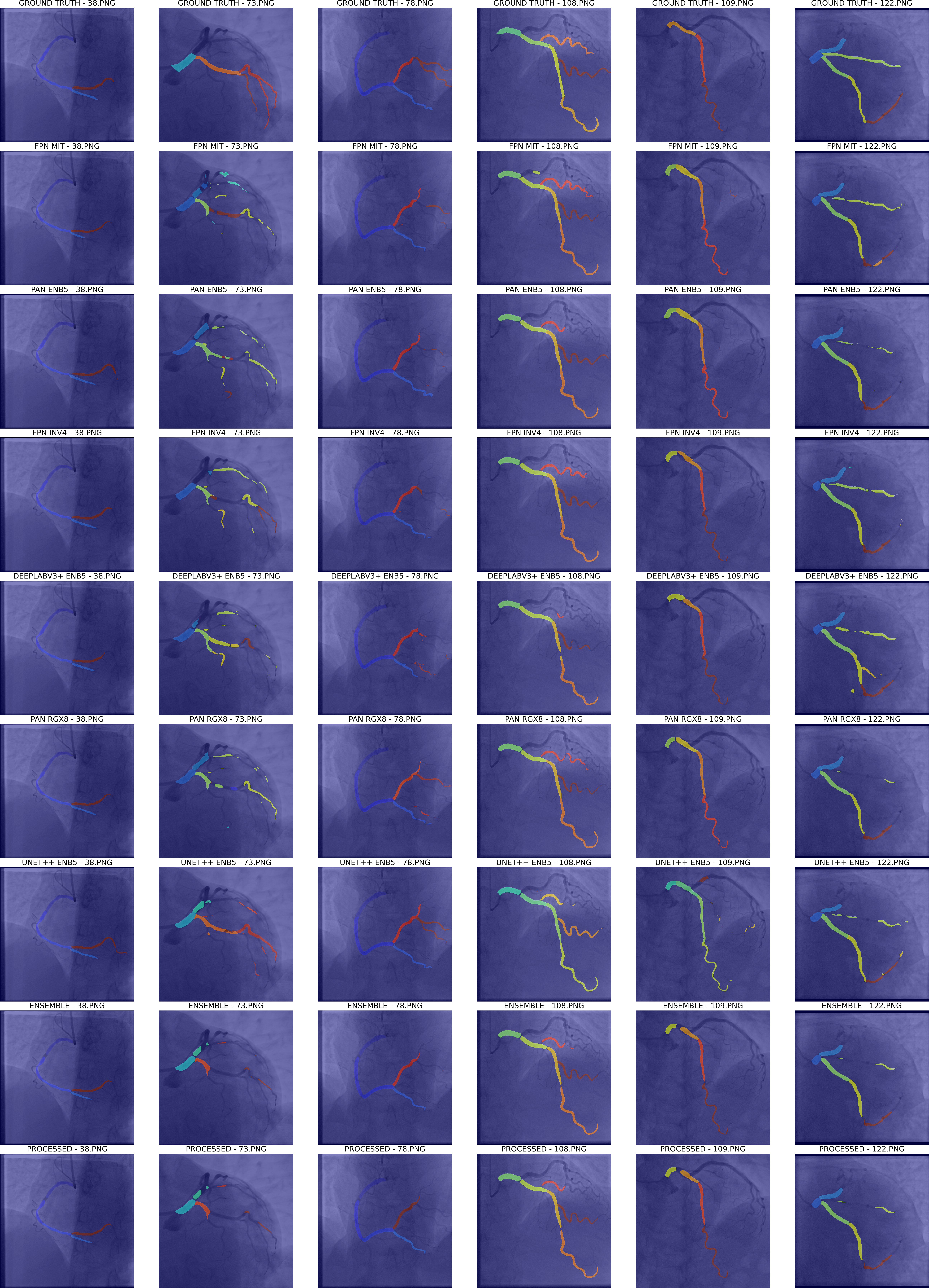}
    \caption{Outputs of Proposed Solution for Comparative Visual Analysis in Multivessel Coronary Artery Segmentation}
    \label{fig:out_fig}
\end{figure}

In summary, our ensemble methodology coupled with robust post-processing, consistently outperformed the baseline models at every training stage. Notably, the baseline models struggled to surpass specific performance thresholds (barely exceeding mean F1 score of 20\%), highlighting their inherent limitations in handling the intricate tasks of coronary artery segmentation and stenosis localisation. Our weighted ensembling approach, capitalising on the strengths of multiple models, achieved substantial performance enhancements, surpassing the average performance of baseline models by more than twofold in both evaluation phases. The solution's performance is further boosted by an additional fold through post-processing refinements. To provide a visual demonstration, Fig. \ref{fig:out_fig} showcases example predictions from the baseline models, the ensemble, and the images after post-processing. It is evident that the segmentation quality significantly improves from the baseline models to the ensemble and post-processed outputs. This study underscores the efficacy of ensemble learning followed by robust post-processing as a pivotal approach in medical AI applications, establishing it as a superior and innovative methodology.

\subsection{Unsuccessful Attempts}
We made several attempts to enhance our solution's performance, but these efforts did not yield the expected results. Some of the unsuccessful approaches we explored include:

\begin{enumerate}
    \item \textbf{Data Preprocessing:} We attempted to enhance contrast in XCA imagery using CLAHE during data preprocessing, but this approach did not produce the desired improvements.
    
    \item \textbf{Geometric Augmentations:} We experimented with geometric augmentations such as Shift Scale rotation, Perspective Warp, and Grid Distortion, but they did not lead to improved results for either of the tasks.
    
    \item \textbf{Synthetic Data Generation:} We considered generating synthetic data, particularly for rare classes like class 10a, using CycleGAN. Unfortunately, we were unable to generate realistic synthetic images.
    
    \item \textbf{Loss Functions:} We explored the use of Dice Loss in combination with Tversky Loss, but our models did not converge as efficiently as they did with Focal Loss and Tversky Loss.
    
    \item \textbf{Fine-Tuning with Phase 1 Validation Data:} Fine-tuning our models with the phase 1 validation set, after its release, did not significantly improve our training.
    
    \item \textbf{Ensembling Strategies:} We experimented with using U-Net with an attention module for ensembling, but this approach did not yield the desired results.
    
    \item \textbf{Conditional Generative Adversarial Networks (CGAN):} We explored the use of CGAN for predicting defect correction, but these efforts did not lead to successful outcomes.
\end{enumerate}

These attempts highlight the challenges and complexities involved in addressing the tasks of coronary artery segmentation and stenosis localization in XCA imagery, and despite our best efforts, certain approaches did not prove effective in achieving our goals.

\section{Limitations and Future Work} 
\label{sec:limits}

While our solution represents a significant advancement, it has certain limitations that require attention for practical applications. Improvements in input preprocessing are needed to balance the noise-to-image ratio, enabling baseline models to acquire robust imaging features and thereby enhancing overall performance. Additionally, while we employed state-of-the-art computer vision architectures and encoders, we acknowledge the potential of generative AI models for medical image segmentation. The iterative refinement offered by diffusion models holds promise for revolutionizing similar tasks, and we plan to explore their integration into our solution.

Furthermore, we envision future research directions in XCA analytics, leveraging emerging visual computing techniques. One exciting prospect involves translating the 2D angiogram representation of the coronary artery tree into 3D, enabling immersive mixed reality applications. This innovation could facilitate collaborative XCA analytics in an immersive environment, improving assessment quality through remote peer guidance during preoperative cardiac surgery planning sessions. Such advancements hold the potential to greatly benefit cardiac care and enhance patient outcomes.

\section{Conclusions}
\label{sec:concs}

In this research, we have proposed a solution for multivessel coronary artery segmentation and stenosis localisation from XCA images. Our solution is based on baseline models, complemented by ensemble techniques for result consolidation, and innovative post-processing to create a robust end-to-end prediction system. While the baseline models successfully learned to segment various aspects of coronary arteries, their individual performance could not surpass certain performance thresholds. This underscores their inherent weaknesses in handling the complexities of coronary artery segmentation and stenosis localisation. The proposed ensemble strategy, with its ability to leverage the strengths of multiple models, yielded significantly improved results, establishing itself as a superior and innovative approach for XCA analytics. Our solution was developed for the ARCADE challenge, where we achieved the 5th position on both leaderboards. We obtained a mean F1 score of $37.69\%$ for coronary artery segmentation and $39.41\%$ for stenosis localisation, showcasing the potential of intelligent tools in assisting CAD diagnosis. These tools not only enhance the reliability of XCA analysis but also hold promise in guiding interventions and improving the accuracy of stent injections in clinical settings. Our work contributes to advancing the field, offering valuable insights and methodologies for future research in this critical area of cardiovascular healthcare.
\bibliographystyle{splncs04}

\begin{thebibliography}{10}
\providecommand{\url}[1]{\texttt{#1}}
\providecommand{\urlprefix}{URL }
\providecommand{\doi}[1]{https://doi.org/#1}

\bibitem{chen2018encoder}
Chen, L.C., Zhu, Y., Papandreou, G., Schroff, F., Adam, H.: Encoder-decoder
  with atrous separable convolution for semantic image segmentation. In:
  Proceedings of the European conference on computer vision (ECCV). pp.
  801--818 (2018)

\bibitem{danilov2021real}
Danilov, V.V., Klyshnikov, K.Y., Gerget, O.M., Kutikhin, A.G., Ganyukov, V.I.,
  Frangi, A.F., Ovcharenko, E.A.: Real-time coronary artery stenosis detection
  based on modern neural networks. Scientific reports  \textbf{11}(1), ~7582
  (2021)

\bibitem{gao2022vessel}
Gao, Z., Wang, L., Soroushmehr, R., Wood, A., Gryak, J., Nallamothu, B.,
  Najarian, K.: Vessel segmentation for x-ray coronary angiography using
  ensemble methods with deep learning and filter-based features. BMC Medical
  Imaging  \textbf{22}(1), ~10 (2022)

\bibitem{kirillov2017unified}
Kirillov, A., He, K., Girshick, R., Doll{\'a}r, P.: A unified architecture for
  instance and semantic segmentation. In: CVPR (2017)

\bibitem{li2018pyramid}
Li, H., Xiong, P., An, J., Wang, L.: Pyramid attention network for semantic
  segmentation. arXiv preprint arXiv:1805.10180  (2018)

\bibitem{nobre2023coronary}
Nobre~Menezes, M., Silva, J.L., Silva, B., Rodrigues, T., Guerreiro, C.,
  Guedes, J.P., Santos, M.O., Oliveira, A.L., Pinto, F.J.: Coronary x-ray
  angiography segmentation using artificial intelligence: a multicentric
  validation study of a deep learning model. The international journal of
  cardiovascular imaging pp. 1--12 (2023)

\bibitem{tao2022lightweight}
Tao, X., Dang, H., Zhou, X., Xu, X., Xiong, D.: A lightweight network for
  accurate coronary artery segmentation using x-ray angiograms. Frontiers in
  Public Health  \textbf{10},  892418 (2022)

\bibitem{tmenova2019cyclegan}
Tmenova, O., Martin, R., Duong, L.: Cyclegan for style transfer in x-ray
  angiography. International journal of computer assisted radiology and surgery
   \textbf{14},  1785--1794 (2019)

\bibitem{zhou2018unet++}
Zhou, Z., Rahman~Siddiquee, M.M., Tajbakhsh, N., Liang, J.: Unet++: A nested
  u-net architecture for medical image segmentation. In: Deep Learning in
  Medical Image Analysis and Multimodal Learning for Clinical Decision Support:
  4th International Workshop, DLMIA 2018, and 8th International Workshop,
  ML-CDS 2018, Held in Conjunction with MICCAI 2018, Granada, Spain, September
  20, 2018, Proceedings 4. pp. 3--11. Springer (2018)

\bibitem{zir1976interobserver}
Zir, L.M., Miller, S.W., Dinsmore, R.E., Gilbert, J., Harthorne, J.:
  Interobserver variability in coronary angiography. Circulation
  \textbf{53}(4),  627--632 (1976)

\end{thebibliography}

\end{document}